\documentclass[12pt]{article}
\usepackage[margin=1in]{geometry}
\usepackage[english]{babel}
\usepackage[T1]{fontenc}
\usepackage{mathpazo}
\usepackage{microtype}
\usepackage{setspace}
\usepackage{amsmath,amssymb,amsfonts,bm,amsthm}
\usepackage{booktabs,threeparttable,tabularx,array}
\usepackage{graphicx,float,caption}
\usepackage{xcolor}
\usepackage{natbib}
\usepackage{hyperref}
\usepackage[autostyle,english=american]{csquotes}
\MakeOuterQuote{"}

\definecolor{medium-blue}{rgb}{0,0,0.5}
\definecolor{ChadBlue}{rgb}{.1,.1,.5}
\definecolor{ChadRed}{rgb}{.5,0,.5}
\hypersetup{colorlinks,linkcolor=ChadRed,citecolor=ChadBlue,urlcolor=medium-blue}

\newtheorem{proposition}{Proposition}
\newtheorem{lemma}{Lemma}
\newtheorem{assumption}{Assumption}
\newtheorem{definition}{Definition}

\newcommand{\mc}{\mathcal}
\newcolumntype{Y}{>{\raggedright\arraybackslash}X}
\title{How Firms Export:\\ An Assignment Theory of Export Organization\thanks{I thank Ignacio Berasategui, Pedro Bom, Francisco Requena, Martin Stojanovik, and participants at the XXVII Conference on International Economics for helpful comments. This research is part of grant PID2025-168794NB-I00, funded by MICIU/AEI/10.13039/501100011033 and by the ESF+. I also gratefully acknowledge the financial support from the Department of Science, Universities and Innovation of the Basque Government (IT1793-26).}}

\author{\large Asier Minondo\thanks{Deusto Business School, University of Deusto, Camino de Mundaiz 50, 20012 Donostia--San Sebasti\'{a}n, Spain. Email: \href{mailto:aminondo@deusto.es}{aminondo@deusto.es}.}}

\date{This version: \today}

\begin{document}
\maketitle

\begin{abstract}
\begin{singlespace}
This paper develops a model in which direct exporters, indirect exporters, pure intermediaries, and hybrid exporters emerge endogenously from the assignment of products to export platforms. Firms differ in manufacturing and commercial capabilities. Manufacturing capability determines a product's potential profitability, while commercial capability determines how effectively a platform can sell it abroad. Manufacturing activation and product assignments are jointly determined, subject to fixed manufacturing costs, contracting frictions, and increasing organizational costs. Complementarity between manufacturing and commercial capabilities generates positive assortative matching, while organizational diseconomies shape platform scope. Equilibrium allocations can feature manufacturers exporting through their own platforms or those of other firms, specialized intermediaries exporting products they do not manufacture, and hybrid firms exporting their own products alongside those manufactured by other firms. The model characterizes how contracting costs, foreign-market attractiveness, and organizational diseconomies affect equilibrium assignments.
\end{singlespace}
\end{abstract}

\begin{flushleft}
\end{flushleft}

\newpage

\section{Introduction}
\label{sec:introduction}

Some manufacturers export their own products directly \citep{bernard2007comparative}. Others reach foreign markets
through wholesalers or other intermediaries \citep{ahn2011intermediaries}. A third group combines production
and intermediation, exporting both their own products and those sourced from other producers
\citep{bernard2019carry,erbahar2023tradeintermediation}. This paper develops a model of product assignment and export platform scope that accommodates these organizational choices.

The central feature of the model is the endogenous allocation of products across the organizations that commercialize them in foreign markets. I refer to these organizations as
export platforms. A platform can be a manufacturer's own commercial division, an independent wholesaler, or the commercial division of another manufacturing firm. Platforms differ in commercial capability: their ability to find buyers, access distribution channels, negotiate, and convert foreign demand into sales. Firms also differ in potential manufacturing capability, which determines marginal production cost if they activate production.\footnote{This separation of capabilities is in the spirit of \citet{hallak2013product}, who model firm heterogeneity in process and product productivity.} Activating a product requires a fixed manufacturing cost, so firms below a capability threshold do not manufacture but may still operate export platforms. Each platform can handle several products, but expanding its portfolio entails increasing organizational costs.

I model the allocation of products to export platforms as a many-to-one assignment market. A producer and a platform can write a contract that coordinates the foreign-market price and divides the resulting match surplus. Each exported product is assigned exclusively to one platform, whereas a platform can acquire several product rights. Convex organizational costs are represented by increasingly costly platform slots. This converts the portfolio problem into an assignment game between products and slots. The formulation makes entry, matching, and scope parts of a single equilibrium allocation.

The model yields four main results. First, match surplus is strictly supermodular in manufacturing and commercial capability. Among arm's-length relationships, more manufacturing-capable producers are therefore assigned to more commercially capable platforms. The complementarity between manufacturing and commercial capability generates positive assortative matching in equilibrium.

Second, commercial capability and product scope are linked in equilibrium. If two platforms face the same organizational cost schedule, a less capable platform cannot operate many more products than a more capable one. 

Third, manufacturing activation, own-product retention, and the acquisition of external products jointly determine export organization. The model accommodates direct exporters, indirect exporters, pure intermediaries, and hybrid firms. I provide a sufficient condition under which these four forms exhaust the set of active firms.

Fourth, the equilibrium responds differently to trade and organizational costs. A fall in the cost of arm's-length contracting weakly increases the number of intermediated relationships. Stronger scope diseconomies weakly reduce the number of product pairs handled by the same platform. 
An increase in foreign demand or a reduction in variable trade costs weakly increases total export operating profits, but need not increase the number of exported products or active platforms.

My contribution is to develop a model that jointly explains
how manufacturers export and why specialized intermediaries
emerge. Manufacturers can export their own products directly,
export through another firm's platform, or combine exports
of their own products with those sourced from other producers.
Firms can also operate as pure intermediaries, commercializing
other producers' products without manufacturing a product
of their own. These organizational forms emerge from
manufacturing and commercial capabilities, the costs of
production and platform operation, and the allocation
of products across platforms.

The model brings together three strands of the literature.
Research on heterogeneous firms explains why some firms
export while others serve only the domestic market
\citep{bernard2007comparative}. Models of trade intermediation
explain the role of specialized intermediaries and why
manufacturers choose between direct and indirect exporting
\citep{ahn2011intermediaries,akerman2018wholesalers}.
The literature on carry-along trade examines why manufacturers
also export products made by other firms
\citep{bernard2019carry,erbahar2023tradeintermediation}.
I bring export participation, specialized intermediation,
and hybrid exporting into a common framework. Manufacturing
activation, platform operation, and product assignment are
determined jointly, allowing direct exporters, indirect
exporters, pure intermediaries, and hybrid firms to arise
within the same equilibrium allocation.

The rest of the paper is organized as follows. Section~\ref{sec:environment} presents the environment and derives match surplus. Section~\ref{sec:equilibrium} defines the assignment equilibrium. Section~\ref{sec:sorting} characterizes sorting and platform scope. Section~\ref{sec:modes} discusses export modes. Section~\ref{sec:comparative} presents comparative statics, and Section~\ref{sec:conclusion} concludes. 

\section{Environment}
\label{sec:environment}

\subsection{The home economy and foreign demand}

There is a small home economy populated by a representative household and a finite set of firms $\mc N=\{1,\ldots,N\}$. Labor is the only factor. The household supplies $L$ units of labor inelastically and owns all home firms. A freely traded numeraire is produced one-for-one with labor. I focus on equilibria in which the numeraire is produced, so its price and the home wage are both one. The numeraire sector absorbs labor not used in production, commercialization, or platform organization. Profits and net export receipts accrue to the household. 

I focus on one foreign destination, indexed by $x$. I abstract from domestic sales of differentiated products and focus on production for export to destination $x$. Thus, a potential differentiated product is manufactured only if it is exported. This assumption keeps the analysis centered on export organization and the assignment of products to commercial platforms. Each home firm $i$ is endowed with the capability to manufacture one potential differentiated product and with the capability to operate an export platform. Activating production of the differentiated product requires a fixed manufacturing cost $F^M>0$. A potential product that is not activated is not manufactured; the firm may nevertheless operate an export platform. Demand in destination $x$ for an active product $i$ at consumer price $p$ is
\begin{equation}
	q_x(p)=D_xp^{-\sigma},
	\qquad D_x>0,\quad \sigma>1.
	\label{eq:demand}
\end{equation}

The demand shifter $D_x$ captures the size of destination $x$. Demand is additively separable across products, so the demand for one product does not depend on the prices or sales of other varieties.

Firm $i$ has potential manufacturing capability $\varphi_i>0$. Conditional on activating production, its marginal production cost is $1/\varphi_i$. The same firm also has commercial capability $\gamma_i>0$ and can operate an export platform regardless of whether it manufactures. A platform is the set of commercial resources used to reach foreign customers. If product $i$ is sold through the platform of firm $j$, the delivered marginal cost is
\begin{equation}
c_{ijx}=\frac{\tau_x d(\gamma_j)}{\varphi_i},
\qquad d(\gamma)=1+\frac{1}{\gamma},
\qquad \tau_x>1.
\label{eq:marginal_cost}
\end{equation}
Thus, manufacturing capability reduces production cost, while commercial capability reduces the variable resources needed to serve the destination.

The analysis does not require manufacturing and commercial capabilities to be independently distributed. The equilibrium results apply to any given configuration of firm capabilities.

\subsection{Coordinated contracts and match surplus}

Producer $i$ and platform $j$ can sign an exclusive agreement
to export product $i$ to destination $x$. The agreement
specifies the final price and a payment between the parties
that divides their joint profit. Under this agreement, the producer and platform choose the
final price $p$ to maximize their joint operating profit:

\begin{equation}
\max_p \left(p-c_{ijx}\right)D_xp^{-\sigma}.
\end{equation}
The optimal price and joint operating profit are
\begin{align}
p_{ijx}&=\frac{\sigma}{\sigma-1}c_{ijx},
\label{eq:price}\\
s_{ijx}&=A_x z_i a_j,
\label{eq:match_surplus}
\end{align}
where
\begin{equation}
A_x\equiv \frac{D_x}{\sigma}
\left(\frac{\sigma-1}{\sigma\tau_x}\right)^{\sigma-1},
\qquad
z_i\equiv\varphi_i^{\sigma-1},
\qquad
a_j\equiv d(\gamma_j)^{1-\sigma}.
\label{eq:transformed_types}
\end{equation}

Higher manufacturing capability implies a higher $z_i$,
while higher commercial capability implies a higher $a_j$. The scalar $A_x$ summarizes market attractiveness: it increases with foreign demand and falls with variable trade costs.

The fixed manufacturing cost does not affect the optimal price, but it determines whether a potential product is activated. Let $a^{\max}=\max_{j\in\mc N}a_j$ denote the commercial
capability of the best available platform. Even on this
platform, product $i$ generates operating profit of at most
$A_xz_i a^{\max}$ before relationship and organizational
costs. If this amount is smaller than the fixed manufacturing
cost $F^M$, the product cannot be profitably manufactured.
A necessary condition for manufacturing is therefore
$A_xz_i a^{\max}\ge F^M$. Hence every firm with
\begin{equation}
z_i<\underline z_x^M\equiv\frac{F^M}{A_xa^{\max}}
\label{eq:manufacturing_cutoff}
\end{equation}
does not manufacture a differentiated product. The corresponding threshold in manufacturing capability is $\underline\varphi_x^M=(\underline z_x^M)^{1/(\sigma-1)}$.

An arm's-length relationship, $i\ne j$, costs $f_x\ge0$. This is the labor required to search, negotiate, adapt information systems, and enforce an interfirm agreement. Placing firm $j$'s own product on its own platform does not require this cost. I refer to the resulting relationship-cost saving as the own-product advantage. 

Subtracting manufacturing and relationship costs from
operating profit gives
\begin{equation}
	b_{ijx}=s_{ijx}-F^M-f_x\mathbf{1}\{i\ne j\}.
	\label{eq:net_pair_surplus}
\end{equation}

\subsection{Platform slots and organizational costs}

Running an export platform involves organizational costs, such as managing distribution, warehousing, and regulatory compliance. Following \citet{akerman2018wholesalers}, I assume that
the incremental organizational cost weakly increases
with the number of products handled, limiting the
incentive to concentrate all products on a single platform.

Let $n_j\in\{0,1,\ldots,N\}$ denote the number of products
handled by platform $j$. Its organizational cost is
$K_x(n_j)$, with $K_x(0)=0$. The incremental cost of adding
the $k$th product is
\begin{equation}
	\Delta K_x(k)=K_x(k)-K_x(k-1).
	\label{eq:incremental_cost}
\end{equation}

\begin{assumption}[Increasing organizational costs]
	\label{ass:convex_cost}
	For every $k$, $\Delta K_x(k)>0$ and
	$\Delta K_x(k+1)\ge\Delta K_x(k)$.
\end{assumption}

I represent these incremental costs by treating each
platform $j$ as having slots $(j,1),\ldots,(j,N)$, where
slot $(j,k)$ accommodates one product at organizational
cost $\Delta K_x(k)$. Assigning product $i$ to this slot
generates joint net surplus for the producer and platform:
\begin{equation}
	v_{i,jk}=b_{ijx}-\Delta K_x(k).
\end{equation}

\section{Assignment equilibrium}
\label{sec:equilibrium}

An allocation $\mu$ determines which products are manufactured
and assigns each of them to one platform slot, with each slot
accommodating at most one product.

If $\mu(i)=(j,k)$, firm $i$ manufactures its product and exports
it through slot $k$ on platform $j$. An unmatched product is
not manufactured, and slots may remain unused.
Section~\ref{sec:environment} established that a product must
be capable of covering its fixed manufacturing cost.
This condition is necessary but not sufficient for production:
relationship and organizational costs, together with competing
uses of platform capacity, also affect the allocation.

Manufacturing and platform operation are separate decisions.
A firm may operate a platform without manufacturing its own
product, or manufacture a product that is exported through
another firm's platform.

Producers and platform slots bargain over surplus. A payoff vector consists of producer payoffs $u_i\ge0$ and slot payoffs $r_{jk}\ge0$. The slot payoffs belonging to firm $j$ accrue to that firm's platform.

\begin{definition}[Assignment equilibrium]
An assignment equilibrium is an allocation $\mu$ and payoffs $(u,r)$ such that:
\begin{enumerate}
\item for every matched pair $\mu(i)=(j,k)$, $u_i+r_{jk}=v_{i,jk}$;
\item unmatched products and slots receive zero; and
\item no producer and platform slot can obtain a larger
combined payoff by matching with each other:
$u_i+r_{jk}\ge v_{i,jk}$ for all $i$ and $(j,k)$.
\end{enumerate}
\end{definition}

This is a standard assignment game \citep{shapley1971assignment}.
Payments between producers and platforms divide the surplus
generated by their matches. The equilibrium can be interpreted
as platforms bidding for the exclusive right to export each
product, with each producer accepting the offer that gives
it the highest payoff.

\begin{proposition}[Existence and efficiency]
\label{prop:existence}
An assignment equilibrium exists. An allocation is supportable as an assignment equilibrium if and only if it solves

\begin{equation}
	\max_{\mu\ \mathrm{feasible}} V_x(\mu),
	\qquad
	V_x(\mu)=
	\sum_{(i,j,k):\,\mu(i)=(j,k)}
	\left[
	A_xz_ia_j-F^M
	-f_x\mathbf{1}\{i\ne j\}
	-\Delta K_x(k)
	\right].
	\label{eq:assignment_problem}
\end{equation}
Equivalently, equilibrium maximizes total export operating
profit net of manufacturing, relationship, and organizational
costs.
\end{proposition}

\noindent\textit{Proof.} See Appendix~\ref{app:proof_existence}.

The platform-formation decision is endogenous. Firm $j$ operates no platform when all its slots are unused, operates a one-product platform when only its first slot is used, and expands until no further product can cover the relevant incremental cost and equilibrium product rent. Equation~\eqref{eq:assignment_problem} jointly determines which platforms operate, the composition of their portfolios, and the allocation of products across platforms.

\section{Sorting and platform scope}
\label{sec:sorting}

\subsection{Positive assortative matching}

Consider two products with $z_h>z_l$ and two platforms with $a_H>a_L$. For any fixed pair of slots, the cross-difference in gross match surplus is
\begin{align}
&A_xz_ha_H+A_xz_la_L-A_xz_ha_L-A_xz_la_H
\nonumber\\
&\hspace{2cm}=A_x(z_h-z_l)(a_H-a_L)>0.
\label{eq:supermodularity}
\end{align}
Thus, manufacturing and commercial capability are strict complements.

\begin{proposition}[Assortative matching]
\label{prop:pam}
Consider two arm's-length matches in equilibrium. If $z_h>z_l$, the product with type $z_h$ cannot be assigned to a platform with lower commercial capability than the platform serving $z_l$. Hence arm's-length product--platform assignment is positive assortative in manufacturing and commercial capability.
\end{proposition}

\noindent\textit{Proof.} See Appendix~\ref{app:proof_pam}.

The result holds for many-to-one matching because the organizational cost belongs to a slot, not to the identity of the product occupying it. Swapping two products across fixed slots leaves organizational costs unchanged, cannot increase relationship costs, and raises operating surplus by Equation~\eqref{eq:supermodularity}. The own-product advantage can create a local exception: a platform may retain its own product even when a slightly better external product would otherwise occupy the slot. Proposition~\ref{prop:pam} therefore applies exactly to arm's-length matches and describes sorting among the products for which matching is empirically meaningful.

\subsection{Commercial capability and product scope}

Section 2.3 introduced organizational costs that rise as
a platform handles more products. These costs limit the
gains from expanding a portfolio. Commercial capability
works in the opposite direction: a more capable platform
generates greater operating profit from any given product
and can therefore justify handling additional products.
This subsection examines how these forces determine
differences in platform size.

Let $n_j(\mu)$ denote the number of products handled by
platform $j$ in allocation $\mu$, which I call its product
scope. Recall that $f_x$ is the relationship cost incurred
when a platform handles another firm's product. A platform
does not incur this cost when handling its owner's product.

When platforms face the same organizational costs and
$f_x=0$, a more commercially capable platform must handle
at least as many products as a less capable platform.
Otherwise, moving a product from the larger, less capable
platform to an unused slot on the more capable platform
would raise operating profit without increasing
organizational costs.

When $f_x>0$, handling its owner's product gives a platform
a cost advantage. Moving that product to another platform
would introduce a relationship cost. This advantage allows
a less commercially capable platform to handle one more
product than a more capable platform, but no more than one.

\begin{proposition}[Commercial capability and platform scope]
	\label{prop:scope}
	Suppose all platforms face the same incremental
	organizational cost schedule, and consider two platforms
	with commercial capabilities $a_H>a_L$.
	\begin{enumerate}
		\item If handling another firm's product entails no
		relationship cost ($f_x=0$), then $n_H\ge n_L$.
		\item If handling another firm's product entails a positive
		relationship cost ($f_x>0$), then $n_H\ge n_L-1$.
		\item If platform $L$ does not handle its owner's product,
		then $n_H\ge n_L$, even when $f_x>0$.
	\end{enumerate}
\end{proposition}

\noindent\textit{Proof.}
See Appendix~\ref{app:proof_scope}.

The result compares platforms within the same equilibrium,
allowing product assignment and platform operation to be
determined jointly. Under a common organizational cost
schedule, greater commercial capability is associated
with weakly larger portfolios, subject to the one-product
exception created by the saving on relationship costs.

\section{Export modes}
\label{sec:modes}

Export modes emerge endogenously from the equilibrium assignment of products to export platforms.

\begin{definition}[Export modes]
For destination $x$:
\begin{enumerate}
\item firm $j$ is a \emph{direct exporter} if its platform carries its own product and no external product;
\item firm $j$ is a \emph{hybrid exporter} if its platform carries its own product and at least one external product;
\item firm $j$ is a \emph{pure intermediary} if it does not manufacture a product and its platform carries at least one product manufactured by another firm; and
\item firm $i$ is an \emph{indirect exporter} if it manufactures a product that is carried by platform $j\ne i$ and its own platform is inactive.
\end{enumerate}
\end{definition}

The first three categories classify active platform operators; the fourth covers manufacturers that export without operating a platform. However, an additional configuration is possible: a firm may export its own product through another firm's platform while operating its own platform to commercialize products manufactured by other firms. I call this configuration \emph{cross-hauling}. Appendix~\ref{app:four_type_partition} provides a sufficient condition that rules out cross-hauling, ensuring that the four export modes form a mutually exclusive and exhaustive partition of exporting firms.

The possibility of cross-hauling illustrates that a firm operating an export platform need not use it to export its own product. More generally, whether a firm retains its own product on its platform depends on the value of the alternatives. The product could instead be exported through another firm's platform or not manufactured at all. Moreover, using a slot for its own product may prevent the platform from carrying another firm's product.

To formalize this choice, the relevant alternatives must allow other products to be reassigned. Let $R_j^{-j}\geq0$ denote the increase in maximized total surplus from allowing product $j$ to be exported through another firm's platform or left unmanufactured, relative to the allocation problem without product $j$. Let $C_{jk}\geq0$ denote the reduction in the maximized surplus of the other products when slot $(j,k)$ is reserved for product $j$. All remaining products can be reassigned when computing these values. Precise definitions are provided in Appendix~\ref{app:proof_own_product}.

Exporting its own product through slot $k$ generates operating profit $A_xz_ja_j$, from which the fixed manufacturing cost $F^M$ and the incremental organizational cost $\Delta K_x(k)$ must be deducted. Since the firm uses its own platform, it avoids the arm's-length relationship cost $f_x$. The remaining surplus must cover both the opportunity cost of reserving the slot and the value of the product's alternative assignments.

\begin{lemma}[Own-product retention]
\label{lem:own_product}
There exists an assignment equilibrium in which firm $j$'s product is carried by its own platform if and only if there is a slot $k$ such that
\begin{equation}
A_xz_ja_j-F^M-\Delta K_x(k)-C_{jk}\geq R_j^{-j}.
\label{eq:own_retention}
\end{equation}
If the inequality is strict for at least one slot, every equilibrium assigns the product to its own platform. If it fails for every slot, no equilibrium does so. If the highest left-hand side equals $R_j^{-j}$, equilibria with and without own-product retention both exist.
\end{lemma}

\noindent\textit{Proof.} See Appendix~\ref{app:proof_own_product}.

Equation~\eqref{eq:own_retention} compares the contribution of own-platform exporting with the value of the alternatives, allowing the rest of the allocation to adjust. Holding the opportunity costs $R_j^{-j}$ and $C_{jk}$ fixed, higher manufacturing or commercial capability makes own-product retention more attractive. This is a conditional comparison: the opportunity costs can themselves change when capabilities change, so the inequality does not establish an unconditional manufacturing-capability cutoff. Firms whose manufacturing capability is too low to justify producing their own product may nevertheless operate as pure intermediaries if operating their platforms is profitable.

These results do not imply that manufacturing and commercial capabilities alone determine a firm's export mode. Because product assignments and their opportunity costs are endogenous, a firm's organizational form also depends on the capabilities of other firms.

\section{Comparative statics}
\label{sec:comparative}

The equilibrium objective can be written compactly as
\begin{equation}
V_x(\mu;A_x,f_x,K_x)
=A_xB(\mu)-F^MP(\mu)-f_xM(\mu)-C(\mu),
\label{eq:compact_objective}
\end{equation}
where
\begin{align}
B(\mu)&=\sum_{i:\mu(i)=(j,k)}z_ia_j,\\
P(\mu)&=\sum_{i:\mu(i)=(j,k)}1,\\
M(\mu)&=\sum_{i:\mu(i)=(j,k)}\mathbf 1\{i\ne j\},\\
C(\mu)&=\sum_{(j,k)\ \mathrm{used}}\Delta K_x(k).
\end{align}
$B$ is the aggregate capability index of active matches, $P$ is the number of manufactured products, $M$ is the number of arm's-length relationships, and $C$ is total organizational cost.

\begin{proposition}[Arm's-length contracting costs]
	\label{prop:relationship_cost}
	Holding all other primitives fixed, let $\mu$ and $\mu'$ be equilibrium allocations at $f_x$ and $f_x'>f_x$, respectively. Then $M(\mu')\leq M(\mu)$: an increase in the cost of arm's-length contracting weakly reduces the number of products exported through another firm's platform.
\end{proposition}

\noindent\textit{Proof.} See Appendix~\ref{app:proof_relationship_cost}.

This result allows products to be reassigned and platforms
to enter, exit, or change scope. Higher contracting costs
can lead a manufacturer to use its own platform or cease
exporting, while other products and platforms adjust.

\begin{proposition}[Foreign demand and variable trade costs]
	\label{prop:market_attractiveness}
	Holding all other primitives fixed, let $\mu$ and $\mu'$ be equilibrium allocations at $A_x$ and $A_x'>A_x$, respectively. Then $B(\mu')\geq B(\mu)$: an increase in foreign demand or a reduction in variable trade costs weakly raises the aggregate capability index of active export matches.
\end{proposition}

\noindent\textit{Proof.} See Appendix~\ref{app:proof_market_attractiveness}.

To examine the effects of stronger scope diseconomies, suppose that the incremental organizational cost of handling the $k$th product is

\[
\Delta K_x(k;\kappa)=\Delta K_x^0(k)+\kappa(k-1),
\]

where $\kappa\geq 0$. 
An increase in $\kappa$ makes larger product portfolios progressively more costly relative to smaller ones.

\begin{proposition}[Scope diseconomies]
	\label{prop:scope_cost}
	Holding all other primitives fixed, an increase in $\kappa$ weakly reduces the total number of pairs of products handled by the same export platform in equilibrium.
\end{proposition}

\noindent\textit{Proof.} See Appendix~\ref{app:proof_scope_cost}.

The result indicates that stronger scope diseconomies reduce an aggregate measure of portfolio concentration. This adjustment can involve reallocating products across platforms, changes in the number of exported products, or both. It does not require every platform to reduce its scope.

\section{Conclusion}
\label{sec:conclusion}

This paper develops a model in which four commonly observed export modes---direct exporting, indirect exporting, pure intermediation, and hybrid exporting---emerge endogenously. Firms differ in manufacturing and commercial capabilities, and products are assigned to export platforms. Manufacturing capability, together with fixed manufacturing costs and the available assignment opportunities, determines which products are produced, while commercial capability determines the value a platform can generate from exporting them. Complementarity between these capabilities generates positive assortative matching, while increasing organizational costs shape the number of products handled by each platform.

These forces jointly determine whether manufacturers export through their own platforms or those of other firms, whether commercially capable firms specialize in intermediating products manufactured elsewhere, and whether manufacturers use their platforms to export their own products alongside those manufactured by other firms. The model thus explains different export modes through a common assignment mechanism rather than separate exporting technologies. It also characterizes relative platform scope, identifies conditions under which the four modes form an exhaustive classification, and establishes how foreign-market attractiveness, contracting costs, and organizational diseconomies affect equilibrium assignments. The central insight is that how firms export depends not only on their own capabilities but also on the capabilities of other firms and the opportunities created by matching products to export platforms.

A natural direction for future research is to extend the assignment framework to multiple destinations. Such an extension would allow manufacturing activation and organizational costs to be shared across markets, potentially generating additional patterns of export organization.

\clearpage
\appendix
\numberwithin{equation}{section}

\section{Proofs}

\subsection{Proof of Proposition~\ref{prop:existence}}
\label{app:proof_existence}

There are finitely many products and slots, so the set of feasible matchings is finite and the maximization problem in Equation~\eqref{eq:assignment_problem} has a solution. To establish that a maximizing allocation can be supported as an assignment equilibrium, express the allocation as indicators: $x_{ijk}=1$ if $\mu(i)=(j,k)$ and $x_{ijk}=0$ otherwise. Relaxing integrality gives the linear program
\begin{align}
\max_{\{x_{ijk}\}}\quad &\sum_{i,j,k}v_{i,jk}x_{ijk}
\label{eq:primal}\\
\text{subject to}\quad
&\sum_{j,k}x_{ijk}\le1 &&\forall i,\nonumber\\
&\sum_i x_{ijk}\le1 &&\forall (j,k),\nonumber\\
&x_{ijk}\ge0 &&\forall i,j,k.\nonumber
\end{align}
The first constraint assigns each product to at most one slot, and the second assigns at most one product to each slot. This is the maximum-weight bipartite matching problem. Its constraint matrix is totally unimodular, so an optimal extreme point is integral and corresponds to a feasible matching. The linear program therefore has the same optimal value as Equation~\eqref{eq:assignment_problem}.

The dual program is
\begin{align}
\min_{\{u_i,r_{jk}\}}\quad &\sum_i u_i+\sum_{j,k}r_{jk}
\label{eq:dual}\\
\text{subject to}\quad
&u_i+r_{jk}\ge v_{i,jk} &&\forall i,j,k,\nonumber\\
&u_i\ge0 &&\forall i,\nonumber\\
&r_{jk}\ge0 &&\forall (j,k).\nonumber
\end{align}
The dual variables are producer and slot payoffs. Strong duality equates the optimal values of the two programs. For any surplus-maximizing allocation and an optimal dual solution, complementary slackness gives $u_i+r_{jk}=v_{i,jk}$ for matched pairs and zero payoffs for unmatched products and slots. Dual feasibility gives $u_i+r_{jk}\ge v_{i,jk}$ for every possible pair. These are precisely the assignment-equilibrium conditions, so every surplus-maximizing allocation can be supported as an equilibrium.

Conversely, consider any assignment equilibrium. For any alternative feasible matching, summing its pairs' inequalities $u_i+r_{jk}\ge v_{i,jk}$ bounds its total surplus by the sum of all equilibrium payoffs, since payoffs are nonnegative and no product or slot appears more than once. That sum equals the surplus of the equilibrium allocation, because matched pairs divide their surplus and unmatched products and slots receive zero. No alternative matching therefore generates more surplus. \qed

\subsection{Proof of Proposition~\ref{prop:pam}}
\label{app:proof_pam}

Suppose product $h$ with $z_h>z_l$ is assigned to a platform with $a_L$, while product $l$ is assigned to a platform with $a_H>a_L$, and both relationships are arm's length. Keep the occupied slots fixed and swap the two products. The swap leaves organizational costs unchanged and cannot increase relationship costs: either new relationship may use the producer's own platform. Manufacturing costs are unchanged. The increase in operating surplus is
\begin{equation}
A_x(z_h-z_l)(a_H-a_L)>0.
\end{equation}
The resulting increase in total surplus is at least this large, contradicting Proposition~\ref{prop:existence}. Therefore no such crossing can occur. \qed

\subsection{Proof of Proposition~\ref{prop:scope}}
\label{app:proof_scope}

Let platforms $H$ and $L$ satisfy $a_H>a_L$, and represent each platform as using its first $n_j$ slots. This entails no loss of surplus because incremental organizational costs are nondecreasing.

First, suppose $f_x=0$. If $n_L>n_H$, choose a product $i$ occupying a slot $(L,k)$ with $k>n_H$. Slot $(H,k)$ is unused. Moving the product to $(H,k)$ leaves manufacturing and organizational costs unchanged and raises operating surplus by $A_xz_i(a_H-a_L)>0$. Relationship costs remain zero, contradicting surplus maximization. Thus $n_H\geq n_L$.

Next, suppose $f_x>0$ and $n_L>n_H+1$. At least two products occupy slots on $L$ with indices greater than $n_H$. Since at most one can be firm $L$'s own product, at least one such slot $(L,k)$ carries an externally sourced product $i\ne L$. Moving it to the unused slot $(H,k)$ preserves manufacturing and organizational costs and raises operating surplus. Relationship costs cannot increase: they remain $f_x$ if $i\ne H$ and fall to zero if $i=H$. The move therefore raises total surplus, a contradiction. Hence $n_H\geq n_L-1$.

Finally, if $L$ does not carry its own product, any occupied slot on $L$ with index greater than $n_H$ carries an externally sourced product. The same move rules out $n_L>n_H$, establishing $n_H\geq n_L$ even when $f_x>0$. \qed

\subsection{Proof of Lemma~\ref{lem:own_product}}
\label{app:proof_own_product}

Let $W_{-j}$ be the maximum total surplus when product $j$ is excluded but all platform slots, including those owned by firm $j$, remain available. Let $W_{-j,-(j,k)}$ be the corresponding maximum when slot $(j,k)$ is also unavailable. Define
\begin{equation}
C_{jk}=W_{-j}-W_{-j,-(j,k)}\geq0.
\label{eq:slot_opportunity_cost}
\end{equation}
Both problems optimize over all feasible assignments of the remaining products. Thus $C_{jk}$ accounts for any reallocation induced by reserving the slot, rather than merely the value of its current occupant.

Let $W_j^{\mathrm{out}}$ be the maximum total surplus when product $j$ may be assigned to another firm's platform or left unmatched, but may not use any slot on its own platform. Define
\begin{equation}
R_j^{-j}=W_j^{\mathrm{out}}-W_{-j}\geq0.
\label{eq:outside_opportunity_value}
\end{equation}
Nonnegativity follows because leaving product $j$ unmatched permits every allocation feasible in the problem defining $W_{-j}$. All other products are freely reassigned in the problem defining $W_j^{\mathrm{out}}$.

If product $j$ is required to occupy slot $(j,k)$, its match contributes $v_{j,jk}=A_xz_ja_j-F^M-\Delta K_x(k)$. The remaining products and slots generate maximized surplus $W_{-j,-(j,k)}$. Hence the highest total surplus attainable with own-product retention is
\begin{align*}
W_j^{\mathrm{own}}
&=\max_k\left\{v_{j,jk}+W_{-j,-(j,k)}\right\}\\
&=W_{-j}+\max_k\left\{A_xz_ja_j-F^M-\Delta K_x(k)-C_{jk}\right\}.
\end{align*}
Every feasible allocation either retains product $j$ on its own platform or satisfies the restriction defining $W_j^{\mathrm{out}}$. The unrestricted maximum is therefore $\max\{W_j^{\mathrm{own}},W_j^{\mathrm{out}}\}$. By Proposition~\ref{prop:existence}, an equilibrium with own-product retention exists if and only if $W_j^{\mathrm{own}}\geq W_j^{\mathrm{out}}$. Subtracting $W_{-j}$ gives Equation~\eqref{eq:own_retention} for at least one slot.

If $W_j^{\mathrm{own}}>W_j^{\mathrm{out}}$, every maximizing allocation retains the product. If the inequality is reversed, no maximizing allocation retains it. If the two values are equal, each restricted problem attains the unrestricted maximum, so Proposition~\ref{prop:existence} supports equilibria both with and without own-product retention. \qed

\subsection{When the four types form a partition}
\label{app:four_type_partition}

Cross-hauling occurs when a firm exports its own product through another firm's platform while operating a platform that handles external products. This subsection establishes a sufficient condition ruling out cross-hauling, ensuring that the four export modes form a mutually exclusive and exhaustive partition of exporting firms. Consider a candidate cross-hauling allocation in which product $j$ is assigned to platform $\ell$ and product $i$ is assigned to platform $j$, with $i\ne j\ne\ell$. Swapping the two products places $j$ on its own platform and sends $i$ to platform $\ell$. The swap saves at least one arm's-length relationship cost. Define the organizational-coherence condition
\begin{equation}
f_x>
A_x\left[(z_j-z_i)(a_\ell-a_j)\right]_+,
\qquad [y]_+\equiv\max\{y,0\},
\label{eq:no_cross_hauling}
\end{equation}
for every feasible cross-hauling pair. The right-hand side is the largest complementarity loss created by replacing the two cross-matches with an own-product match.

\begin{proposition}[Four-type partition]
\label{prop:partition}
If Condition~\eqref{eq:no_cross_hauling} holds, no equilibrium contains cross-hauling. Consequently, every exporting firm belongs to exactly one of four mutually exclusive categories: direct exporter, indirect exporter, pure intermediary, or hybrid firm.
\end{proposition}

\noindent\textit{Proof.} Suppose, contrary to the claim, that product $j$ is exported through platform $\ell\ne j$ while platform $j$ carries product $i\ne j$. Keep the two occupied slots fixed and swap the products. Before the swap, both matches are arm's length and pay $2f_x$. After the swap, product $j$ uses its owner's platform and product $i$ pays at most one relationship cost, so the swap saves at least $f_x$. Organizational costs are unchanged. The change in operating surplus is
\begin{equation}
A_x(z_j-z_i)(a_j-a_\ell)
=-A_x(z_j-z_i)(a_\ell-a_j).
\end{equation}
Condition~\eqref{eq:no_cross_hauling} implies that the relationship-cost saving strictly exceeds any positive operating-surplus loss. The manufacturing costs are unchanged by the swap. The swap therefore raises equilibrium value, contradicting Proposition~\ref{prop:existence}. Cross-hauling cannot occur. An active platform operator must consequently be direct, hybrid, or pure; a manufacturer exporting without an active platform is indirect. These categories are mutually exclusive. \qed

\subsection{Proof of Proposition~\ref{prop:relationship_cost}}
\label{app:proof_relationship_cost}

Optimality at $f_x$ and $f_x'$ implies
\begin{align}
A_xB(\mu)-F^MP(\mu)-f_xM(\mu)-C(\mu)
&\ge A_xB(\mu')-F^MP(\mu')-f_xM(\mu')-C(\mu'),\\
A_xB(\mu')-F^MP(\mu')-f_x'M(\mu')-C(\mu')
&\ge A_xB(\mu)-F^MP(\mu)-f_x'M(\mu)-C(\mu).
\end{align}
Adding and simplifying gives $(f_x'-f_x)[M(\mu)-M(\mu')]\ge0$. Since $f_x'>f_x$, $M(\mu')\le M(\mu)$. \qed

\subsection{Proof of Proposition~\ref{prop:market_attractiveness}}
\label{app:proof_market_attractiveness}

The two revealed-preference inequalities are
\begin{align}
A_xB(\mu)-F^MP(\mu)-f_xM(\mu)-C(\mu)
&\ge A_xB(\mu')-F^MP(\mu')-f_xM(\mu')-C(\mu'),\\
A_x'B(\mu')-F^MP(\mu')-f_xM(\mu')-C(\mu')
&\ge A_x'B(\mu)-F^MP(\mu)-f_xM(\mu)-C(\mu).
\end{align}
Adding gives $(A_x'-A_x)[B(\mu')-B(\mu)]\ge0$. Hence $B(\mu')\ge B(\mu)$. \qed

\subsection{Proof of Proposition~\ref{prop:scope_cost}}
\label{app:proof_scope_cost}

Let $n_j(\mu)$ denote the number of products assigned to platform $j$ under allocation $\mu$. Because incremental organizational costs are nondecreasing, an equilibrium allocation can be represented with each active platform using its first $n_j(\mu)$ slots.

Under the assumed cost specification, the additional organizational cost associated with $\kappa$ is

\[
\kappa\sum_j\sum_{k=1}^{n_j(\mu)}(k-1)
=\kappa\sum_j\binom{n_j(\mu)}{2}.
\]

Define $S(\mu)=\sum_j\binom{n_j(\mu)}{2}$, and let $Q(\mu)$ denote total surplus net of manufacturing, relationship, and baseline organizational costs. The equilibrium objective can then be written as

\[
V_x(\mu;\kappa)=Q(\mu)-\kappa S(\mu).
\]

Let $\mu$ and $\mu'$ be equilibrium allocations at $\kappa$ and $\kappa'>\kappa$, respectively. Optimality implies

\begin{align*}
	Q(\mu)-\kappa S(\mu)
	&\geq Q(\mu')-\kappa S(\mu'),\\
	Q(\mu')-\kappa' S(\mu')
	&\geq Q(\mu)-\kappa' S(\mu).
\end{align*}

Adding these inequalities and rearranging yields

\[
(\kappa'-\kappa)[S(\mu)-S(\mu')]\geq 0.
\]

Since $\kappa'>\kappa$, it follows that

\[
\sum_j\binom{n_j(\mu')}{2}
\leq
\sum_j\binom{n_j(\mu)}{2}.
\]

Thus, stronger scope diseconomies weakly reduce the total number of product pairs handled by the same platform. \qed

\section{A parametric organizational-cost schedule}
\label{app:parametric_cost}

A convenient example satisfying Assumption~\ref{ass:convex_cost} is
\begin{equation}
K_x(n)=\kappa_{1x}n+\frac{\kappa_{2x}}{1+\eta_x}n^{1+\eta_x},
\qquad \kappa_{1x}>0,\quad \kappa_{2x}>0,\quad \eta_x>0.
\end{equation}
The linear component captures the setup and compliance resources required for each active product. The convex component captures congestion within the platform. The incremental cost is
\begin{equation}
\Delta K_x(k)=\kappa_{1x}
+\frac{\kappa_{2x}}{1+\eta_x}
\left[k^{1+\eta_x}-(k-1)^{1+\eta_x}\right],
\end{equation}
which is positive and increasing in $k$.

\end{document}